# Quantum Interference Enhances the Performance of Single-Molecule Transistors

Zhixin Chen[1]*, Iain M. Grace[2], Steffen L. Woltering[1,3], Lina Chen[1,3], Alex Gee[1], Jonathan Baugh[4], G. Andrew D. Briggs[1], Lapo Bogani[1], Jan A. Mol[5], Colin J. Lambert[2]*, Harry L. Anderson[3]*, and James O. Thomas[1]*

[1] Department of Materials, University of Oxford, 16 Parks Road, Oxford OX1 3PH, UK

[2] Department of Physics, Lancaster University, Lancaster, LA1 4YB, UK

[3] Department of Chemistry, University of Oxford, Chemistry Research Laboratory, Oxford OX1 3TA, UK

[4] Institute for Quantum Computing, University of Waterloo, N2L 3G1 Waterloo, ON, Canada

[5] School of Physical and Chemical Sciences, Queen Mary University, London, E1 4NS, UK

*zhixin.chen@materials.ox.ac.uk          *c.lambert@lancaster.ac.uk          *harry.anderson@chem.ox.ac.uk
*james.thomas@materials.ac.uk

**An unresolved challenge facing electronics at a few-nm scale is that resistive channels start leaking due to quantum tunneling. This affects the performance of nanoscale transistors, with single-molecule devices displaying particularly low switching ratios and operating frequencies, combined with large subthreshold swings.[1] The usual strategy to mitigate quantum effects has been to increase device complexity, but theory shows that if quantum effects are exploited correctly, they can simultaneously lower energy consumption and boost device performance.[2-6] Here, we demonstrate experimentally how the performance of molecular transistors can be improved when the resistive channel contains two destructively-interfering waves. We use a zinc-porphyrin coupled to graphene electrodes in a three-terminal transistor device to demonstrate a >$10^4$ conductance-switching ratio, a subthreshold swing at the thermionic limit, a > 7 kHz operating frequency, and stability over >$10^5$ cycles. This performance is competitive with the best nanoelectronic transistors. We fully map the antiresonance interference features in conductance, reproduce the behaviour by density functional theory calculations, and trace back this high performance to the coupling between molecular orbitals and graphene edge states. These results demonstrate how the quantum nature of electron transmission at the nanoscale can enhance, rather than degrade, device performance, and highlight directions for future development of miniaturised electronics.**

Tunneling field-effect transistors (TFETs)[7] and single-molecule transistors (SMTs)[8] are devices where quantum effects in electron transmission, normally considered detrimental to the performance of transistors with nanometre dimensions,[9] become responsible for the function of the device. Using a single molecule as an active channel brings the benefit of synthesis with atomic precision, providing the possibility to control quantum effects through molecular design to enable high performance,[10] as well as leading to complementary functionalities such as thermoelectric recovery of waste heat, multistate switching, or sensing.[11] Quantum interference (QI) is a characteristic quantum effect found in nanoscale charge transport.[12-14] However, it is difficult to create two nanoscale quantum-coherent channels in standard conductors, because scattering and defects quickly lead to loss of electron coherence. Consequently, the practical use of QI has been almost exclusively limited to superconducting devices to obtain extremely sensitive magnetometers,[15] and its potential for transistors remains largely unexplored. Nevertheless, theory predicts that harnessing it is a promising route to high-performance SMTs or efficient thermoelectric generators.[2] To evaluate these predictions requires fundamental experimental investigation into the specific impact of QI on transistor properties.

Destructive QI (DQI) can be controlled by electrochemical gating, whereby conductance switching over two orders of magnitude was achieved for several cycles.[5,6] However, electrochemical gating is relatively slow and incompatible with many practical applications. Graphene source and drain electrodes enable more versatile electrostatic gating and measurement of QI in single-molecule devices.[16] Furthermore, non-trivial transmission effects resulting from the coupling between the graphene density of states with molecular orbitals can sometimes enhance the device properties.[17,18]

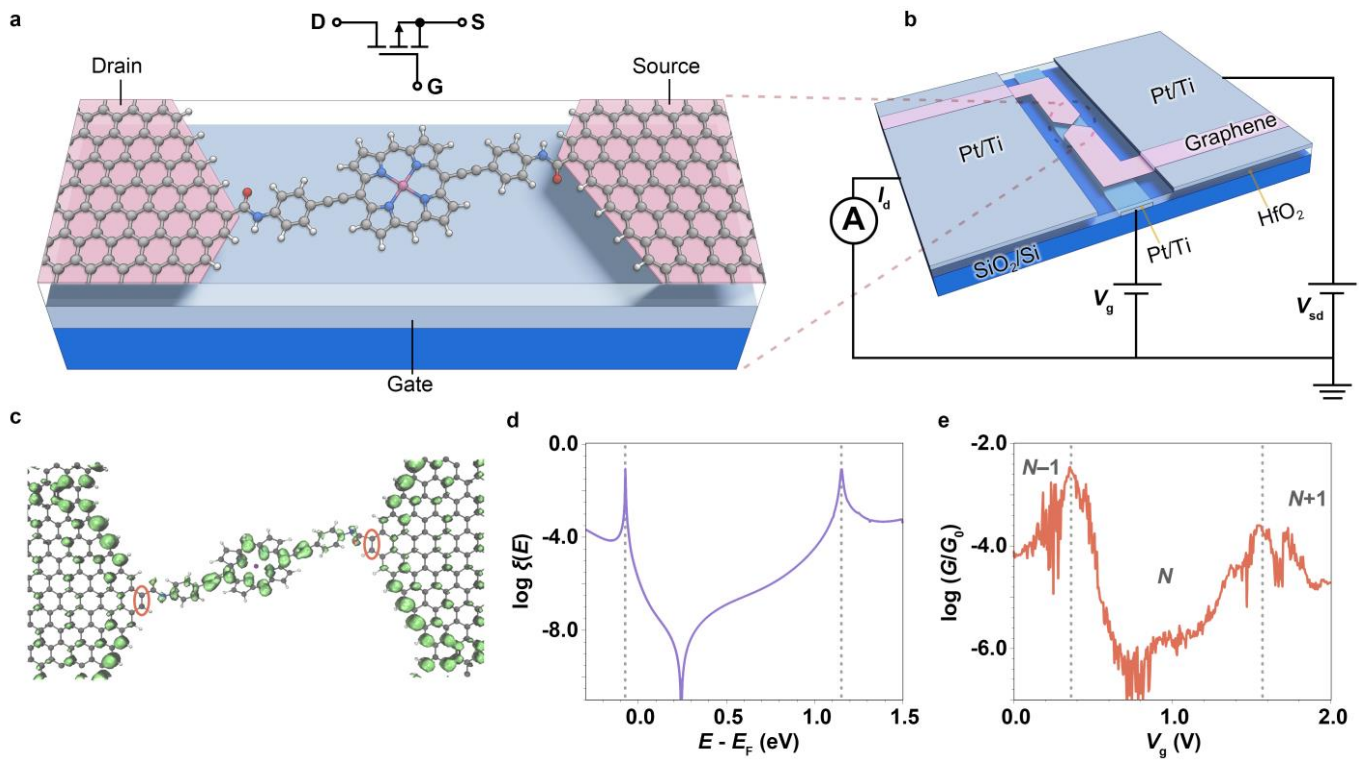

**Fig. 1 | Quantum-interference-enhanced single-molecule transistor. a** Schematic representation of a graphene-based single-molecule transistor. The 3,5-*bis*(trihexylsilyl)phenyl solubilising groups on the lateral *meso* positions of the porphyrin have been replaced with H atoms for clarity. **b** Device architecture. The grey-blue rectangular strip in the centre is the local platinum gate electrode under a 10 nm layer of HfO$_2$ (transparent); the rectangular areas (grey-blue) at each end are source and drain platinum electrodes, which are in contact with the bowtie-shape graphene (pink). **c** Optimised junction geometry with the local density of states at the Fermi level displayed in green (isovalue is set at 0.0005). The zero-LDOS carbon atoms are highlighted in red. **d**. Calculated behaviour of the electronic transmission ξ as a function of energy. **e** Differential conductance at $T$ = 80 K vs $V_g$ at $V_{sd}$ = 0 mV. The conductance is plotted in logarithmic scale as the ratio to conductance quantum, $G_0$.

To explore the use of QI in nanoelectronic devices, we employ a zinc-porphyrin with 4-ethynylaniline anchor groups at opposite (5,15) *meso* positions and bulky 3,5-bis(trihexylsilyl)phenyl substituents at the other two (10, 20) *meso* positions as an active channel (Fig. 1a, see Supplementary Section 2 for synthesis). The energy-level spacings and the chemical potentials are found to be within the experimental ranges of source-drain and gate voltages ($V_{sd}$ and $V_g$, respectively). As a result, both the on- and off- resonance transport regimes are accessible, allowing evaluation of not only the interference between orbitals, but also the influence of their coupling to the reservoirs. The molecules were integrated into three-terminal molecular transistor devices by direct covalent coupling of amine groups to carboxylic acid residues on the oxidised edges of graphene electrodes, which are generated during electroburning (see Methods).[19] The current $I_{sd}$ is measured on applying $V_{sd}$ and the device behaviour can be switched using $V_g$ (Fig. 1b).

The atomically-defined nature of the molecular transport channel allows the prediction of the device behaviour using a combination of DFT and quantum transport theory.[20,21] For the isolated molecule, the highest occupied molecular orbital (HOMO) and the lowest unoccupied one (LUMO) are not predicted to interfere destructively (Supplementary Section 8).[22] However, the molecular coupling to the carbon atoms of the electrodes has a significant influence on transmission: the connection to carbon atoms with zero local density of states at the Fermi energy, as present at irregular graphene edges,[17,23] leads to a dramatic QI effect (Fig. 1c, d). The resulting electron transmission spectrum, ξ(E), spans over ten decades, with an asymmetric shape, and an extremely pronounced dip produced by anti-resonance within the HOMO-LUMO gap (Fig. 1d). This is a surprising result: contrary to the usual case where QI is produced by the phase properties of the frontier molecular orbitals,[22] the anti-resonance here arises as a result of coupling between graphene edge states via molecular orbitals (Supplementary Section 8).

The experimental behaviour displays the predicted features, as shown by plotting the gate-dependence of the zero-bias conductance (normalised by the conductance quantum),[12] $G_{sd} = \partial I_{sd}/\partial V_{sd} \propto$ $\xi(E)$ (Fig. 1e). The peaks in the conductance trace arise from resonant charging of the molecule, and at $V_g$ ~ 0.4 V transport occurs via a change in occupation of the HOMO, while at $V_g$ ~ 1.6 V the LUMO dominates transport, as detailed later in the full single-electron transistor characterization. The asymmetric anti-resonance feature predicted by the calculations is observed in the region between these peaks where the molecule remains neutral, showing a pronounced dip arising from the DQI effect. For a large region around $V_g$ = 0.78 V the conductance at the dip reaches down to below the lowest detection level of our set-up, ~$10^{-7.0}$ $G_0$. This demonstrates that modulating $V_g$ moves the molecular levels of the porphyrin between being on-resonance, where conductance is at local maxima, to off-resonance and complete conductance suppression.

The QI effects can now be used to operate a transistor device. When charging is considered, the full behaviour includes neutral molecular states, with $N$ electrons, and oxidised ($N$–1) and reduced ($N$+1) states produced by varying $V_g$. High ON:OFF current ratios are afforded by using the DQI-induced conductance dip as OFF state, and the $N$–1/$N$ resonant tunneling channel as the ON state (Fig. 2a). The full mapping of the current vs $V_{sd}$ and $V_g$ shows reproducible single-electron-transistor behaviour (Fig. 2b and 2c), and locally steep responses of current to gate voltage which are advantageous for the efficient switching of the device (Fig. 2d). Regions of low current are when the device is in the off-resonant condition where first-order tunneling processes are suppressed by Coulomb blockade, and low residual current is carried by phase-coherent off-resonant transmission. These regions (known as Coulomb diamonds) are separated by the resonant tunneling regions, when a molecular level is within the bias window generated by applying $V_{sd}$, where current is high (Fig. 2b). The number of electrons on the molecule is fixed within each Coulomb diamond, and varies by one between adjacent diamonds,[8] leading to the definitions of the $N$–1, $N$, $N$+1 states mentioned above. The coexistence of Coulomb blockade and phase-coherent off-resonant tunneling demonstrates the device is in the intermediate molecule-electrode coupling ($\Gamma$) regime,[24,25] and from the FWHM of a Coulomb peak we find $\Gamma$ = 8 meV.

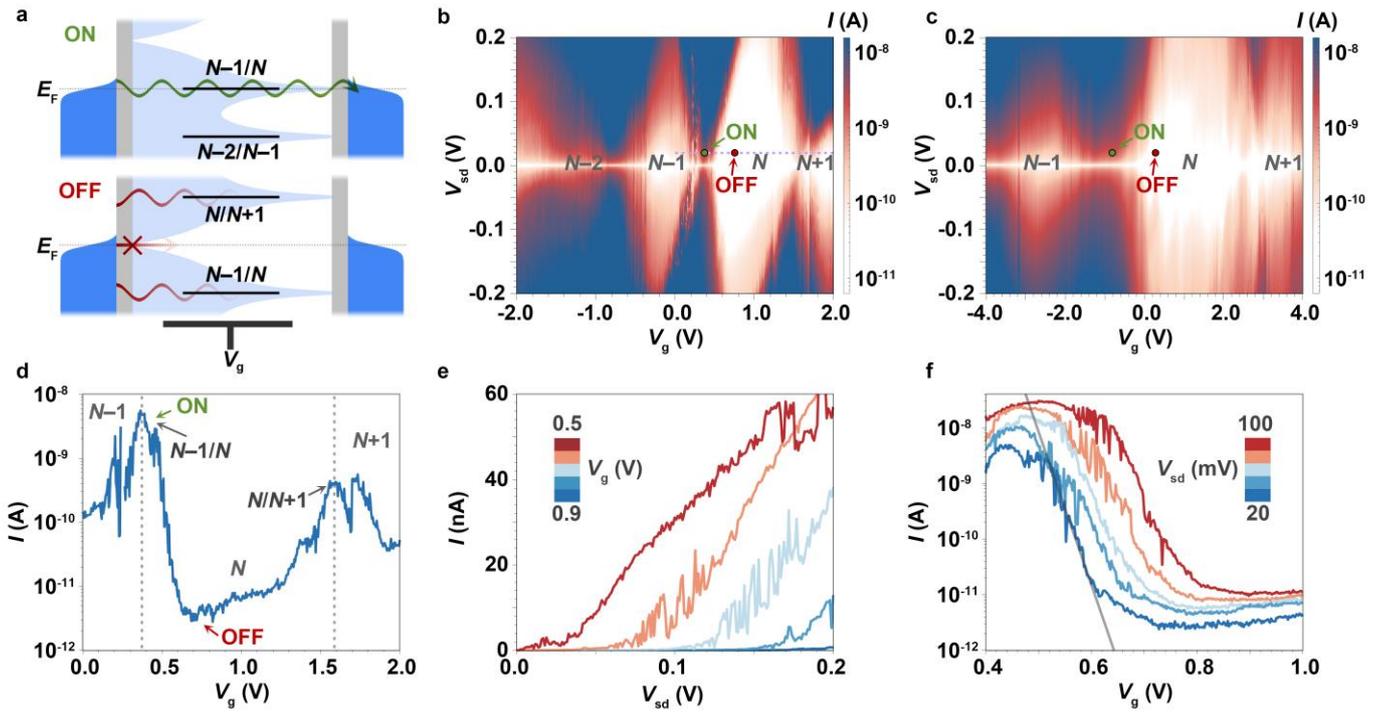

**Fig. 2 | Quantum-interference transistor properties. a** Schematic model of quantum-interference-based transistor, where conductance is high when a molecular level is on resonance, and low when a phase difference between two pathways suppresses transmission in the $N$ state (the HOMO-LUMO gap). **b**, **c** Current map ($I_{sd}$ vs. $V_{sd}$ and $V_g$) for devices 1 and 2 respectively. Device 1 is discussed in detail in main text, device 2 is characterised in the SI (Fig. S3-4). **d** $I_{sd}$-$V_g$ at $V_{sd}$ = 20 mV (along dashed line on panel **b**). **e** Output characteristics and **f** transfer characteristics of the single-porphyrin transistor, the slope of the grey line shows the thermionic limit at the measurement temperature. All measurements were done at 80 K.

In Fig. 2e and 2f we focus on the transistor performance of the device through plots of the output ($I_{sd}$-$V_{sd}$) and transfer characteristics ($I_{sd}$-$V_g$) between the ON and OFF states. Within the dip ($V_g$ = 0.9 V) Coulomb blockade and DQI lead to extremely low currents with a resistance of ~ 1 GΩ, whereas on the $N$–1/$N$ resonance ($V_g$ = 0.5 V) resonant tunneling via the porphyrin HOMO leads to an $I_{sd}$-$V_{sd}$ that is approximately linear with a resistance of ~ 3 MΩ (Fig. 2e). Shifting the gate potential to move the device between ON and OFF states leads to a current ratio of $10^{3-4}$, depending on $V_{sd}$, and despite the differences in switching mechanism between typical MOSFETs and our SMT, the transfer characteristics have a similar shape (Fig. 2f). There is an approximately linear increase in $\log_{10}(I_{sd})$ as $V_g$ is swept from OFF to ON before a saturation of the source-drain current; from these data we calculate the subthreshold swing ($S_{s-th}$) of the single-molecule device. We find a value of $S_{s-th}$ = 14.5 ±0.4 mV/dec at $V_{sd}$ = 20 mV and 80 K, after adjusting for the unoptimised gate coupling parameter, $\alpha_g$. The value is very close to the thermionic limit – the lower limit on the subthreshold swing for a MOSFET (15.9 mV/dec at 80 K) which we attribute to the steepness of the transmission conferred by DQI.

We test the switching frequency limits of the device by applying a square wave to the gate ($V_{g,min}$ = 0.61 V, $V_{g,max}$ = 0.76 V) with a fixed bias voltage of $V_{sd}$ = 100 mV. Robust current switching is observed at kHz frequencies (Fig. 3a, Fig. S4-1). The response of the current to the voltage is rounded by the RC time, $\tau$, of the circuit. In this case $\tau$ = 30 ± 1 μs, giving a rise/fall time (10% to 90% of ON state current for rise time, *vice versa* for fall time) of 2.2 × $\tau_{rise/fall}$ = 66 ± 2 μs, and a maximum switching frequency (10% to 90%) of $1/(\tau_{rise} + \tau_{fall})$ = 7.6 ± 0.3 kHz. The frequency is limited by the bandwidth of the current amplifier, rather than the intrinsic switching mechanism of the molecular device, as the same RC time is found by measuring the output of the circuit with a 100 MΩ resistor in place of the molecular device (Fig. S4-1). The switching mechanism does not involve any specific conformational change of the molecule, just a transition from off-resonant to resonant tunneling. As sequential resonant tunneling is first order in Γ, we estimate the intrinsic switching frequency could be up to $hf$ ~ Γ ~ 8 meV (found from fitting the left side of the resonance peak), around 1 THz. Therefore the upper frequency limit of switching for the actual device will be limited by the time taken to charge the gate, a property that depends on the dimensions and material of the electrode. Optimising the gate specifications is required for future device development but beyond the scope of our study of the effect of DQI on molecular transistor performance.

A map of $G_{sd}$ vs $T$ shows that the on-state resonance width remains constant below 30K, and then broadens on increasing $T$ due to temperature effects on the electrode Fermi distributions (Fig. 3b). We also observe that the conductance on resonance begins to increase above 30 K, suggesting that thermally-activated processes begin to contribute to the conduction mechanism (Fig. 3c, Fig. S7-1). We limit our analysis to 100 K, because of the difficulty of accounting for charge trapping and vibrational effects at graphene edges at higher temperatures.[26,27] The DQI dip, measured at $V_g$ ~ 0.8 V, where the transport mechanism is solely off-resonance phase-coherent transport, remains close to our minimum detection limit in the $T$ = 10–100 K region. The slight positive $T$-dependence is likely the result of a reduced influence of the DQI-induced suppression, which relies on coherent electron transmission, due to dephasing produced at higher temperature by inelastic tunneling and vibrations at the dynamic molecular/graphene interfaces.[28] The positive $T$ dependence of both on and off states lead to stable ratios above 40 K.

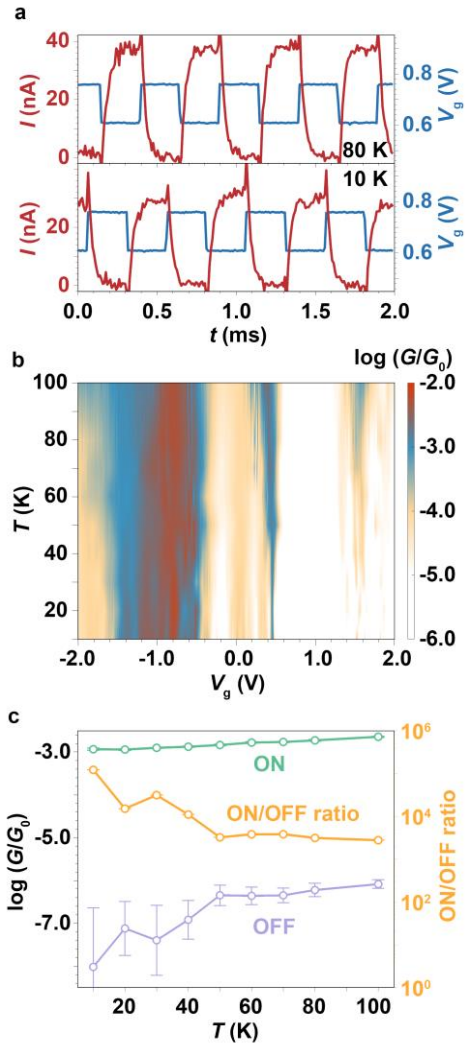

**Fig. 3 | Switching and temperature dependent behaviour at the thermionic limit. a** ON/OFF switching of the device with a 2 kHz square wave applied to the gate at $V_{sd}$ = 100 mV at 80 K (top) and 10 K (bottom). Longer time traces and intermediate temperatures are shown in Fig. S6-1. **b** Differential conductance map measured as a function of temperature and gate voltage at $V_{sd}$ = 0 mV. **c** Conductance for ON state (green), and OFF state (purple), and the conductance ON/OFF ratio (orange) at $V_{sd}$ = 20 mV.

We now examine, through the subthreshold swing, how efficient switching of the device can be achieved. The best gate control that can be achieved in a FET is the thermionic limit that results from the exponential tails of the Fermi distributions of the electrodes (Fig. 4a),[29] however in nanodevices performance is usually degraded further by short-channel parasitic effects. The tunneling current in a single-molecule junction is the convolution of $\xi(E)$ with the difference in electrode Fermi distributions $I_{sd} \propto \int \xi(E)[f_S(\mu_S) - f_d(\mu_d)]dE$, and so a limit on $S_{\text{S-th}}$ related to $T$ will also apply. To understand the effect of DQI on this limit, we compare our experimental measurements of $S_{\text{s-th}}$ at different temperatures to a simulation of $S_{\text{s-th}}$ vs. $T$ around $N$–1/$N$ using the single-level model.[24] The single-level model treats the molecular resonance as a Breit-Wigner resonance, and as it only considers transmission through a single channel it cannot capture interference effects. The lifetime broadening conferred by $\Gamma$ (= 8 meV, the experimentally derived value) and thermal broadening both contribute to the subthreshold swing in a single-molecule transistor, ensuring it remains above the thermionic limit at all $T$, with a small change where $k_B T \ll \Gamma$ that becomes linear as $T$ increases (Fig. 4a). This result reveals a fundamental trade-off when designing a three-terminal nanodevice for transistor applications: a larger $\Gamma$ is desirable to give high ON state currents, but comes at the expense of higher OFF state currents and larger subthreshold swings. Conversely, a smaller $\Gamma$ puts $S_{\text{S-th}}$ closer to the thermionic limit, but reduces tunneling currents to smaller values, leading to high resistances on resonance, thereby limiting ON-OFF ratios. Utilising DQI removes the need to compromise: DQI suppresses off-resonant phase-coherent transport, leading to the steep energy-dependent transmission through the molecular device, even with an intermediate Γ, thereby permitting a high ON-state and low OFF-state current that can be switched by only a small change in $V_g$.

This effect is demonstrated here explicitly in Fig 4b, where DQI accounts for the difference between the simulated and experimental data. At 10 K DQI reduces $S_{s-th}$ from 14.1 mV/dec to 3.6 ± 0.4 mV/dec (thermionic limit: 2.0 mV/dec), at 80 K from 26.3 mV/dec to 14.5 ± 0.4 mV/dec (thermionic limit: 15.9 mV/dec). These observations shows the value of DQI in the device performance: it effectively negates the additional contribution of lifetime broadening to the subthreshold swing, reducing the value to the thermionic limit, even in the intermediate coupling regime.

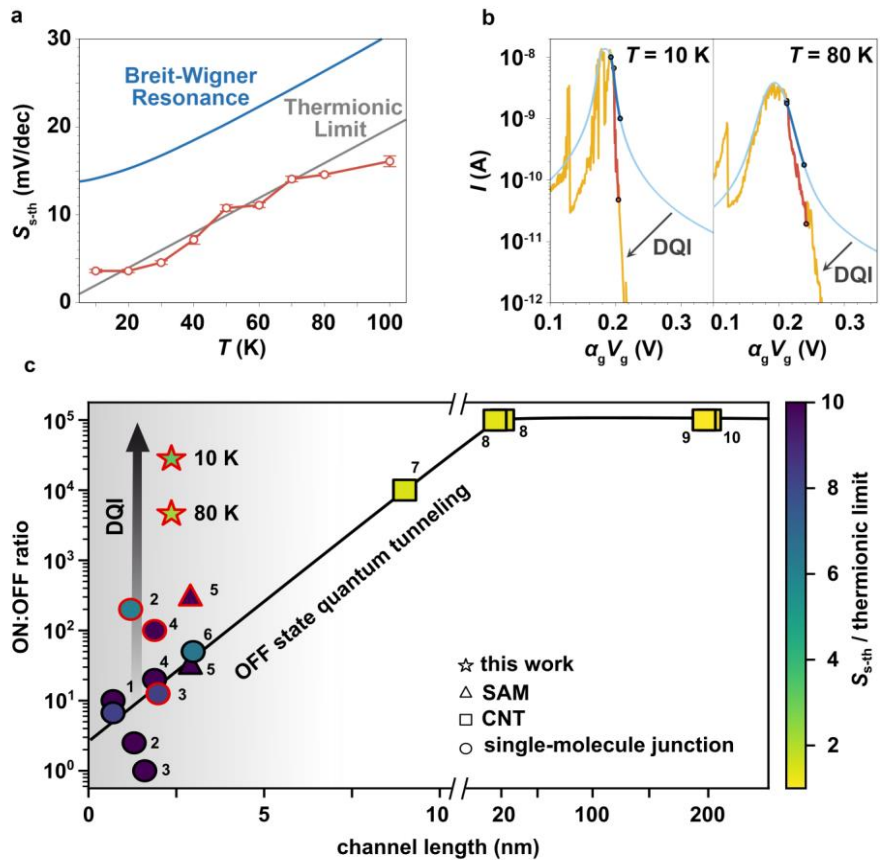

**Fig. 4 | Overview of quantum-Interference-enhanced transistor performance. a** Normalised subthreshold swing at $V_{sd}$ = 20 mV as a function of temperature (red circles, see Fig. S5-1 to Fig. S5-8 for $S_{S-th}$ calculation), plotted with the thermionic limit for a classical FET (grey), and simulated subthreshold swing for a single-level molecular Breit-Wigner resonance without DQI (blue). **b** $I_{sd}$ as a function of $α_g V_g$ around the $N$–1/$N$ resonance, yellow curves are experimental and blue are simulated for a Breit-Wigner resonance. The ranges to calculate subthreshold swing are the steepest parts of the curves, and are highlighted (red for experimental and blue for simulated), and demonstrate the effect of DQI to increase the magnitude of the gradient of the $N$ state current, thereby reducing $S_{S-th}$ to the thermionic limit. **c** Scatter plot of ON:OFF ratio and molecular lengths for a number of molecular and carbon nanotube transistor devices (**1**,[30] **2**,[5] **3**,[31] **4**,[6] **5**,[32] **6**,[33] **7**,[34] **8**,[35] **9**,[36] **10**[37] and this work, with the normalization for $α_g$ removed); SAM = self-assembled mononlayer, CNT = carbon nanotube. Marker colours represent the device subthreshold swing normalised by the thermionic limit at the measurement temperature, which is room temperature except for those specifically marked. Markers with red outlines represent devices that have DQI in their transmission, the black line below 10 nm is a fit to the data points without interference.

The general relationship between ON:OFF ratio and channel length in sub-10 nm transistors (Fig. 4c), can be understood by considering the molecule as a quantum tunneling barrier – even if the device is tuned to the OFF state (i.e. modulating $V_g$ to give a high tunnel barrier), devices inevitably become exponentially more transmissive with decreasing molecular length (i.e. decreasing barrier width), raising OFF currents and decreasing the switching ratio. Without considering the phase-coherent nature of electron transmission, this leakage current fundamentally limits transistors performance on this length scale, shown by the fit in Fig 4c. If devices have DQI in their transmission functions, however, this need not be the case, low OFF currents even with short (< 1nm) molecular lengths,[38] are possible if two out-of-phase coherent transport channels (such as edge states or molecular orbitals) can interfere to suppress overall device transmission. Fig. 4c demonstrates this effect where devices at the scale of a few nm that utilise DQI can have properties of those with longer channel lengths. Utilising this mechanism is a particular advantage of molecular nano-electronics, as the energies and phase properties of orbitals are routinely engineered through molecular design and synthesis. The reduction in ON:OFF ratio with increasing temperature due to increasing leakage current (Fig. 3c, 4c), is attributed to a partial quenching of interference effects,[28] however, a fuller understanding of the underlying mechanism is crucial to design devices that retain extremely low OFF currents resulting from DQI at room temperature. The numbers we report compare favourably to previous studies of SMT properties that have yielded subthreshold swings in the range of 400–1000 mV/dec (our value would be 140 mV/dec at room temperature removing the adjustment for $\alpha_g$ and linearly extrapolating from Fig. 4a). Whilst most SMTs are characterised by a handful of cycles at a frequency below a few Hertz due to inherent timescales of electrochemical gating, we demonstrate a switching frequency of ~7 kHz and over the course of our measurements we switch our device over $10^5$ times. Furthermore, previous measurements are mostly based on STM break-junction techniques so involve continuous reforming of molecular junctions, with the conductance of ON and OFF states taken as an average over many traces, rather than our measurements of repeatedly switching the same, static unimolecular device.

Overall, these results reveal how quantum interference effects can be harnessed in devices just a few nm wide, in pursuit of developing low-power miniaturised electronics. The performances attained offer proof of concept of nanodevices in which quantum effects are used as a resource to enhance device function, rather than being a limitation. Our demonstration makes specific use of the density of states fluctuations at graphene edges, revealing a hitherto undisclosed difference from standard metal electrodes. These concepts, using phase-coherent transmission, can be translated to a series of new compounds and device architectures that are designed to optimally exploit quantum interference. The mild fabrication method allows using a very wide range of chemical compounds to create these nanoscale transistors, opening the path to the creation of multifunctional devices, e.g. with optical or spintronic properties, where interference can be used to control multiple effects at the same time.

**Methods**

*Substrate Fabrication.* On an *n*-doped silicon wafer with a 300-nm-thick $SiO_2$ layer, the gate electrode (3 μm wide) was defined by optical lithography with lift-off resist and electron-beam (e-beam) evaporation of titanium (5 nm) and platinum (15 nm). Next, an ALD-grown dielectric layer of $HfO_2$ (10 nm) was deposited. Finally, source and drain contact electrodes separated by a 7 μm gap (gap centred with the gate) were defined by optical lithography with lift-off resist and e-beam evaporation of titanium (5 nm) and platinum (45 nm).

*Graphene patterning.* PMMA-protected CVD-grown graphene was wet-transferred to the substrate. The PMMA was removed in warm acetone for three hours. The graphene tape with bow-tie shaped structure was patterned by e-beam lithography (EBL) with bi-layer lift-off resist (PMMA495 and PMMA950) and thermal evaporation of aluminium (50 nm). After lift-off, the graphene on unexposed areas (not covered by aluminium) was etched with oxygen plasma. The aluminium was subsequently removed by aqueous NaOH solution (1.0 g in 50 mL water). The sample was finally immersed in warm acetone overnight to remove any residual PMMA. SEM images can be found in Supplementary Section 1.

*Graphene nanogaps.* Graphene nanogaps were prepared by feedback-controlled electroburning of the graphene bow-tie shape until the resistance of the tunnel junction exceeds 1.3 GΩ ($10^{-7}$ $G_0$), see SI for electroburning curves.[39,40] $HfO_2$ has a high dielectric constant and, in combination with the weaker screening of the gate field by using graphene in place of bulky 3D metallic electrodes (most commonly used in molecular junctions),[41] this yields a high gating efficiency. The empty nanogaps were characterised by measuring a current map as a function of bias voltage ($V_{sd}$) and gate voltage ($V_g$) at room temperature in order to exclude devices containing residual graphene quantum dots,[42] only clean devices were selected for further measurement, see Fig. S3-1 and Fig. S3-3 for before and after current maps of the devices presented here.

*Molecule junctions and measurements.* The mechanism of graphene breakdown under electroburning in air is oxidation,[19] and the oxygen-containing functional groups that are thus formed at the edges can be used to engineer the junction by covalent binding.[43-45] Single-molecule devices are fabricated using a condensation reaction between molecular amine groups and carboxylic groups on the edge of the graphene nanogap to form amide bonds.[45] Chips with freshly prepared graphene nanogaps were immersed in 0.5 mL of dry NEt$_3$, after which 20 mg of the amide coupling reagent, hexafluorophosphate azabenzotriazole tetramethyl uronium (HATU), was added. Then, 0.5 mL of a dry $CH_2Cl_2$ solution of **ZnP** (0.2 mmol/L) was added. The mixture was then left at room temperature in the dark for 48–72 h. After the reaction, the chip was washed with $CH_2Cl_2$ and IPA, and blown dry with nitrogen. Then, the chip containing molecular devices was connected to a chip holder via wire bonding, loaded in Oxford Instruments 4K PuckTester, and cooled down to cryogenic temperature for detailed measurements. All current ($I_{sd}$) maps and traces presented in the main text and supplementary files are unprocessed. Conductance data ($G_{sd}$) was calculated by applying a Savitzky–Golay filter (window length: 15, order: 5) to $I_{sd}$ and then differentiating to give: $G_{sd} = \partial I_{sd}/\partial V_{sd}$. Subthreshold swings ($S_{s\text{-th}}$) were calculated from the steepest part of a $\log_{10}(I_{sd})$ vs. $V_g$ trace (over two decades). The measurement was repeated six times for each temperature, see Fig. S5-1 to S5-9 for the traces, and then averaged. The error reports is twice the standard deviation.

*Theoretical Calculations.* Geometry optimizations were carried out using Gaussian 16[46] for the isolated molecular structure using the University of Oxford Advanced Research Computing (ARC) facility, and SIESTA was used for the graphene-based junctions.[47] Transmission spectra were calculated from the Hamiltonian and overlap matrices of the DFT calculation of the junction using GOLLUM[48] quantum transport code. Full details are given in Supplementary Section 8.

**Data Availability**

All the data supporting the findings of this study are available within the Article, its Supplementary Information or from the corresponding authors upon request.

**Acknowledgements**

The authors would like to acknowledge the use of the University of Oxford Advanced Research Computing (ARC) facility in carrying out part of this work (http://dx.doi.org/10.5281/zenodo.22558). The work was supported by the EPSRC (grants EP/N017188/1 and EP/R029229/1), and EU-CoG-MMGNRs. JAM acknowledges funding from the Royal Academy of Engineering and a UKRI Future Leaders Fellowship, Grant No. MR/S032541/1. We thank J. Cremers for help with the synthesis.

**Author Contributions**

The experiments were conceived by J.O.T. and Z.C. with support from G.A.D.B, C.J.L., J.A.M., L.B. and H.L.A.; Z.C. performed the graphene patterning, and undertook the charge-transport measurements with input from A.G.; S.L.W. and J.O.T. synthesised and characterised the compounds under the supervision of H.L.A.; J.B. prepared the device substrates. I.M.G. and Z.C. performed the DFT calculations under the supervision of C.J.L.; J.O.T., L.C,. and Z.C. analysed the data. J.O.T. and Z.C. wrote the paper; all authors discussed the results and edited the manuscript.